# Utilizing Large Language Models for Automating Technical Customer Support

Jochen Wulf[1] and Jürg Meierhofer[1]

[1] Zurich University of Applied Sciences,
Technikumstrasse 81, 8401 Winterthur, Switzerland

**ABSTRACT**

*The use of large language models (LLMs) such as OpenAI's GPT-4 in technical customer support (TCS) has the potential to revolutionize this area. This study examines automated text correction, summarization of customer inquiries and question answering using LLMs. Through prototypes and data analyses, the potential and challenges of integrating LLMs into the TCS will be demonstrated. Our results show promising approaches for improving the efficiency and quality of customer service through LLMs, but also emphasize the need for quality-assured implementation and organizational adjustments in the data ecosystem.*

## 1. INTRODUCTION

Many providers of technical products and services struggle to provide reliable technical customer support (TCS) with fast response times, due to several challenges such as skills shortages and information overload (Özcan et al., 2014). Large Language Models (LLMs) such as OpenAI's GPT-4 have the potential to revolutionize TCS by providing efficient and personalized support ( (Kanbach et al., 2023; Wulf & Meierhofer, 2023) They have the potential to handle a high volume of customer interactions, as well as reduce the need for extensive human resources, and promise significant cost savings (Liu et al., 2023)

Non-generative AI has already transformed customer service, with AI-based tools that automate customer service and sales, perform back-office tasks, and enable remote monitoring, coaching, training, and scheduling (Doellgast et al., 2023). Initial experience in application suggests that machine learning plays a crucial role in automating the resolution of TCS problems and requests, thereby minimizing business interruptions. It increases the efficiency of TCS processes, such as incident and service request management, by automating manual tasks such as categorizing and prioritizing tickets, reducing errors and improving overall process efficiency (Monning et al., 2018).

While initial academic studies are gathering evidence of the business impact of generative AI – and LLMs in particular – the potential of LLMs in transforming customer service is not yet fully understood (Brynjolfsson et al., 2023). In addition, the use of LLMs for TCS creates new challenges regarding data sharing that are largely unknown. The purpose of this paper is to identify cognitive tasks in TCS that can be automated. In addition, this paper evaluates the feasibility with real customer data through prototypes and discusses the organizational challenges regarding data sharing.

## 2. THEORETICAL BACKGROUND
### 2.1 Large Language Models (LLMs)

Large language models have recently demonstrated remarkable capabilities in natural language processing tasks and beyond (Dasgupta et al., 2023; Radford et al., 2019; Wei et al., 2022). These models combine various technical innovations, especially in the domains of architectural design, training strategies, improvements in context length, fine-tuning, multimodality, and scope of training datasets (Liu et al., 2023). However, LLMs face challenges such as hallucination, outdated knowledge, and non-transparent, untraceable processing (Zhao et al., 2023).

Prompt engineering is an increasingly important skill to communicate effectively with LLMs. Prompts are instructions given to an LLM to enforce rules, automate processes, and ensure the quality of the output generated (White et al., 2023). They are a form of programming that can customize the outputs and interactions with an LLM. This includes both System Messages, which are instructions given to the model, and Human Messages, which include the user's inputs into the model (McTear & Ashurkina, 2024).

Retrieval Augmented Generation (RAG) has emerged as a promising solution by incorporating knowledge from external databases (Gao et al., 2023). This improves the accuracy and credibility of the models, especially for knowledge-intensive tasks, and allows for continuous knowledge updates and the integration of domain-specific information. RAG synergistically combines the intrinsic knowledge of LLMs with the extensive, dynamic information in external databases.

## 2.2 The role of AI in technical customer service

Previous research has shown that well-designed customer service practices promote value cocreation (Winkler & Wulf, 2019). Setting up an efficient TCS is not a trivial task. It requires a high degree of know-how in the design and implementation of operational processes as well as in the introduction of the necessary technologies (Wulf & Winkler, 2020).

The academic literature on the application of LLMs for customer service addresses various challenges and industries. Carvalho and Ivanov (2023) outline the profound impact that can be expected on tourism from the integration of ChatGPT and other LLMs, such as improving front-office customer service and the efficiency of back-office operations. Research on customer satisfaction in the public sector points to the value of LLMs in the analysis of online user feedback. Thematic models that can transform user opinions into actionable insights have been proposed to improve the delivery of public services. Studies show that the quality of employee interactions is strongly correlated with user satisfaction (Kowalski et al., 2017).

LLMs are known for their importance in answering questions and chatbots used in healthcare, education, and customer service. For example, developing scalable clustering pipelines to fine-tune LLMs was crucial to identify user intent from large volumes of conversational texts, improve the performance of data analysts, and ultimately reduce the time it takes to deploy chatbots (Chen & Beaver, 2022)

Reinhard et al. (2024) discuss how generative AI can lead to more efficient and high-quality customer support. They identify five support tasks that can be improved with LLMs: ticket capture, ticket assignment, escalation, problem localization, solution customization, solution development, and documentation. Brynjolfsson et al. (2023) empirically examine the introduction of an LLM-based conversational assistant among 5,179 customer service agents and show a 14% increase in productivity, benefiting beginners and low-skilled workers in particular. The AI tool also improved customer sentiment, reduced management intervention requests, and improved employee retention. In summary, existing studies indicate significant potential for the use of LLMs in TCS. However, these studies are largely theoretical in nature. The limited number of prototype and implementation studies shows only a fraction of the potential benefits discussed. The analysis of the literature reveals a research gap in demonstrating the practicality and presenting the necessary conditions for automating cognitive tasks with LLMs in TCS.

## 3. RESEARCH METHODOLOGY

We use prototyping as a research methodology. In design research, prototyping is used to concretize abstract concepts and validate their technical feasibility (Barzilai & Ferraris, 2023; Camburn et al., 2017). Therefore, prototyping is well suited to investigate theoretical approaches for the application of LLMs in TCS that have been discussed in the previous literature.

As a prototyping methodology, we realize software-based proofs of concepts (PoCs) based on data on technical customer requests of a large telecommunications operator. The data sets consist of the textual customer request and an exchange of messages that describe one or more possible solutions. To enable manual validation of PoCs and keep complexity low, we use a random selection of 15 customer requests. Table 1 provides an overview of customer inquiries.

| Inquiry ID | Topic | Number of Words | Number of Chars |
|------------|-------|-----------------|-----------------|
| 1 | Router | 1454 | 9356 |
| 2 | Internet Security | 297 | 2073 |
| 3 | Internet Security | 355 | 2638 |
| 4 | LAN | 1243 | 8423 |
| 5 | Smartphone App | 1266 | 8341 |
| 6 | Router | 777 | 4815 |
| 7 | Router | 1305 | 8310 |
| 8 | WLAN | 557 | 3614 |
| 9 | WLAN | 620 | 3951 |
| 10 | Router | 1076 | 7373 |
| 11 | Router | 555 | 3540 |
| 12 | Connection | 2909 | 17892 |
| 13 | Connection | 2457 | 17190 |
| 14 | Connection | 1485 | 9291 |
| 15 | Contract | 1444 | 9172 |

Table 1: Overview of technical customer inquiries

We implement PoCs for three essential cognitive tasks in TCS: text correction, text summarization, and question answering. In the technical implementation, we use the OpenAI API with the LLMs gpt-4-0125-preview and gpt-3.5-turbo-0125, as these models are optimized for completing tasks and answering questions and represent the current state-of-the-art in the field of LLMs (OpenAI et al., 2023). We also use Chroma DB as a vector database in the RAG architecture and OpenAI's text-embedding-3-small. We use LangChain as the overarching software framework. For validation, we manually compare the LLM outputs with the messages and solutions generated by the human support agents and create quantitative quality metrics, which we discuss in detail below.

## 4. RESULTS
### 4.1 Text Correction
LLMs are able to convert text from one language or language mode to another, as well as correct spelling and grammar errors. They achieve this by understanding the patterns and structures of different languages that they learn from the data they are trained with.

For the PoC, we have written a reply email for each of the 15 customer requests. Then we randomly added classic typos, especially letter twisters and missing letters. Table 2 shows the number of words, characters and spelling errors for the uncorrected reply emails. In the next step, we generated corrected reply mails with the prompt shown in Figure 1 and gpt-3.5-turbo-0125.

```
prompt = ChatPromptTemplate.from_messages([
    ("system", "You are a language expert. Please correct the following
email:"),
    ("user", "{input}")
])
```

Figure 1: Prompt correction of spelling errors in emails

Table 2 shows the number of words, number of characters and the number of remaining errors for the automatically generated corrected reply emails.

| Inquiry | Uncorrected reply mail | | | Corrected reply mail | | |
|---|---|---|---|---|---|---|
| | # Words | # Chars | # Errors | # Words | # Chars | # Errors |
| 1 | 172 | 1173 | 25 | 172 | 1185 | 0 |
| 2 | 153 | 1240 | 15 | 159 | 1268 | 1 |
| 3 | 153 | 1060 | 15 | 155 | 1068 | 0 |
| 4 | 220 | 1573 | 21 | 221 | 1587 | 0 |
| 5 | 198 | 1334 | 24 | 200 | 1348 | 0 |
| 6 | 187 | 1345 | 24 | 187 | 1358 | 0 |
| 7 | 203 | 1607 | 23 | 204 | 1620 | 0 |
| 8 | 143 | 1066 | 16 | 146 | 1082 | 1 |
| 9 | 149 | 1128 | 19 | 154 | 1187 | 2 |
| 10 | 218 | 1585 | 35 | 226 | 1669 | 3 |
| 11 | 208 | 1423 | 31 | 210 | 1439 | 0 |
| 12 | 258 | 1753 | 34 | 260 | 1771 | 0 |
| 13 | 261 | 1827 | 32 | 263 | 1843 | 0 |
| 14 | 138 | 964 | 16 | 143 | 975 | 0 |
| 15 | 205 | 1368 | 22 | 200 | 1378 | 0 |

Table 2: Comparison of uncorrected and corrected emails

A manual analysis of the corrections as well as the quantitative key figures show that LLMs can generate reliable corrections without changing the wording and message content of the error-ridden email. Almost all typos were eliminated by this automated process.

### 4.2 Text Summarization

Modern LLMs build on the transformer architecture and make use of a functionality known as the attention mechanism (Vaswani et al., 2017) This mechanism allows the model to consider the context of the use of each individual word. As a result, the model can understand the relationships between different sections of text and create a summary that accurately reflects the overall meaning of the original text.

In first-level support, it is important to quickly record the communication exchange that has already taken place on a customer request. Automatically generated text summaries could represent a high added value here.

To test this functionality, we had summaries of 100, 200 and 500 words created for each of the 15 data sets consisting of the customer request and the message exchange for solutions. For this purpose, we use the prompt shown in Figure 2 as well as gpt-4-0125-preview.

```
prompt = ChatPromptTemplate.from_messages([
    ("system", "You are a helpful assistant. Summarize the following text in
exactly {num_words}."),
    ("user", "{input}")
])
```

Figure 2: Prompt to generate text summaries

For a systematic comparison of the summary with the source text, we calculate the cosine similarity (1=max, -1=min) of the two corresponding embedding vectors. Cosine similarity is a measure that measures the similarity between two vectors. The results are shown in Table 3. It can be seen that the cosine similarities are between .63 (minimum value) and .86 (maximum value). In addition, it is clearly evident that the cosine similarity increases with longer summaries.

| | | # Words Summary | | | Cosine Similarity | | | Time Saved in Minutes | | |
|---|---|---|---|---|---|---|---|---|---|---|
| Inquiry | # Words | 100 | 200 | 500 | 100 | 200 | 500 | 100 | 200 | 500 |
| Inc1 | 1454 | 115 | 167 | 266 | 0.69 | 0.70 | 0.75 | 5.63 | 5.41 | 4.99 |
| Inc10 | 1076 | 94 | 123 | 239 | 0.63 | 0.63 | 0.72 | 4.13 | 4.00 | 3.52 |
| Inc11 | 555 | 90 | 152 | 216 | 0.76 | 0.80 | 0.84 | 1.95 | 1.69 | 1.42 |
| Inc12 | 2909 | 129 | 190 | 336 | 0.69 | 0.71 | 0.77 | 11.68 | 11.42 | 10.81 |
| Inc13 | 2457 | 117 | 162 | 348 | 0.67 | 0.70 | 0.79 | 9.83 | 9.64 | 8.86 |
| Inc14 | 1485 | 149 | 175 | 267 | 0.69 | 0.67 | 0.68 | 5.61 | 5.50 | 5.12 |
| Inc15 | 1444 | 90 | 162 | 247 | 0.71 | 0.74 | 0.79 | 5.69 | 5.39 | 5.03 |
| Inc2 | 297 | 82 | 137 | 142 | 0.81 | 0.84 | 0.85 | 0.90 | 0.67 | 0.65 |
| Inc3 | 355 | 75 | 115 | 265 | 0.74 | 0.74 | 0.82 | 1.18 | 1.01 | 0.38 |
| Inc4 | 1240 | 92 | 146 | 332 | 0.71 | 0.77 | 0.83 | 4.82 | 4.60 | 3.82 |
| Inc5 | 1266 | 87 | 182 | 253 | 0.75 | 0.80 | 0.80 | 4.95 | 4.55 | 4.26 |
| Inc6 | 777 | 120 | 170 | 307 | 0.81 | 0.82 | 0.86 | 2.76 | 2.55 | 1.97 |
| Inc7 | 1305 | 95 | 146 | 263 | 0.74 | 0.75 | 0.82 | 5.08 | 4.87 | 4.38 |
| Inc8 | 557 | 94 | 146 | 279 | 0.74 | 0.78 | 0.83 | 1.95 | 1.73 | 1.17 |
| Inc9 | 620 | 76 | 161 | 266 | 0.80 | 0.84 | 0.83 | 2.29 | 1.93 | 1.49 |

Table 3: Number of words, cosine similarity and time saved in minutes

In summary, it can be said that gpt-4-0125-preview is able to generate short summaries with a given word count relatively reliably. At the specification of 100 words, the text length average of the summaries produced was 100.33 words. Here, a comparative test with gpt-3.5-turbo-0125 resulted in significant text length deviations. However, with the text length specification of 500 words, gpt-4-0125-preview produced significantly shorter text (average: 268.4 words). A qualitative analysis of the summaries confirms the implication derived from the cosine similarities that the summaries reflect the essential contents of the solution discussion in a relatively robust manner.

Table 3 also shows the estimated time saved when reading the summarized information. The calculation is based on empirical evidence that the average reading speed is 238 words per minute (Brysbaert, 2019). With the approach presented, an average of 4.6 minutes (for 100-word summaries) or 3.9 minutes (for 500-word summaries) of time savings can be achieved per customer request.

### 4.4 Question Answering

When answering questions, the LLM seeks and uses either the internal factual knowledge provided in the pre-training corpus or the external contextual data provided in the prompt to generate answers to questions or instructions. TCS regularly processes customer inquiries about problems that have already been identified. In this processing, historical data can be used to work out solutions to problems. The analysis of historical data can be automated with LLMs.

In a prototype, we generate 10 synthetic customer inquiries, one for each real customer inquiry, each addressing the same problem but using different formulations. First, we then use a vector search (Gao et al., 2023) to search the historical datasets for comparable customer inquiries. In vector search, the prompt is used as a search term and we specify how many similar chunks the search returns. An overview of the search results is shown in Table 4. If only one text chunk is returned per search, the 10 searches will return 100% text chunks that are relevant for problem solving because they are part of the correct data set. With three sections of text, the figure is still 87%.

| Number of chunks returned by vector search | Average proportion of relevant chunks |
| --- | --- |
| 1 | 100% |
| 2 | 95% |
| 3 | 87% |

Table 4: Proportion of relevant text chunks returned by vector searches

In the next step, we embed vector search in a RAG architecture (Gao et al., 2023). To do this, we use the System Message shown in Figure 3 in the prompt. In the context field, the result of the vector search is dynamically inserted. The synthetic customer request is then added to the human message. We send the prompt assembled in this way to gpt-3.5-turbo-0125.

```
SYSTEM_TEMPLATE = """
Answer the user questions in detail and explain all the necessary solution
steps. If the context doesn't contain any relevant information to answer the
question, just say "I don't know":

<context>
{context}
</context>
"""
```

Figure 3: System message for Q&A prompt

In order to validate the content of the LLM answers, we use vector search in addition to qualitative comparisons. We enter the LLM answers into the vector search and extract the cosine distances (0=min, 2=max) to the nearest text chunks of the 15 historical solution descriptions to the customer inquiries in columns Inc1 to Inc10. The results are shown in Table 5.

| | Synthetic Customer Inquiry | | | | | | | | | |
|---|---|---|---|---|---|---|---|---|---|---|
| **Historic Inquiry** | **Inc1** | **Inc2** | **Inc3** | **Inc4** | **Inc5** | **Inc6** | **Inc7** | **Inc8** | **Inc9** | **Inc10** |
| **Inc1** | 0.40 | 0.86 | 1.28 | 1.04 | 1.08 | 0.94 | 0.85 | 1.26 | 1.06 | 0.95 |
| **Inc10** | 0.82 | 0.88 | 1.33 | 0.96 | 1.18 | 1.13 | 1.01 | 1.12 | 0.96 | 0.48 |
| **Inc11** | 1.01 | 1.00 | 1.36 | 1.07 | 1.13 | 1.19 | 0.91 | 1.08 | 1.04 | 1.08 |
| **Inc12** | 0.98 | 0.88 | 1.25 | 0.92 | 1.09 | 1.09 | 0.71 | 1.00 | 1.10 | 0.93 |
| **Inc13** | 0.85 | 0.92 | 1.29 | 0.98 | 1.11 | 1.11 | 0.89 | 1.19 | 1.09 | 1.11 |
| **Inc14** | 1.10 | 0.99 | 1.30 | 1.21 | 1.16 | 1.22 | 0.72 | 0.77 | 1.21 | 1.23 |
| **Inc15** | 1.06 | 1.11 | 1.41 | 1.17 | 1.20 | 1.13 | 0.99 | 1.10 | 1.20 | 1.21 |
| **Inc3** | 1.16 | 0.86 | 0.69 | 1.16 | 1.18 | 1.37 | 0.92 | 1.22 | 1.16 | 1.19 |
| **Inc4** | 0.89 | 0.95 | 1.30 | 0.77 | 1.12 | 1.03 | 0.99 | 1.21 | 0.94 | 0.99 |
| **Inc5** | 1.06 | 0.86 | 1.32 | 1.26 | 0.46 | 1.29 | 1.01 | 1.34 | 1.33 | 1.31 |
| **Inc6** | 0.93 | 1.25 | 1.36 | 1.07 | 1.24 | 0.44 | 1.04 | 1.27 | 1.10 | 1.04 |
| **Inc7** | 0.93 | 0.94 | 1.21 | 0.93 | 1.10 | 0.97 | 0.37 | 1.09 | 1.04 | 1.01 |
| **Inc8** | 1.14 | 1.09 | 1.47 | 1.11 | 1.20 | 1.19 | 0.96 | 0.46 | 1.20 | 1.10 |
| **Inc9** | 1.06 | 1.01 | 1.37 | 0.90 | 1.19 | 1.13 | 1.15 | 1.02 | 0.60 | 0.96 |
| **Inc2** | 1.05 | 0.35 | 1.24 | 1.07 | 0.91 | 1.28 | 0.89 | 1.32 | 1.17 | 1.20 |

Table 5: Cosine Distances of the 10 LLM Responses to the Nearest Text Chunks of the 15 Historical Customer Inquiries

The results show that the cosine distances to the relevant historical solution descriptions are minimal in each case (marked in dark green, mean value 0.5). In addition, it can be seen that the distances to the most similar irrelevant text chunks are considerably larger (mean value 0.9). Inc4 is an exception. There, the cosine distance to the relevant chunk is high at 0.77 and the cosine distance to the nearest irrelevant chunk is relatively low at 0.9. However, a qualitative analysis of the synthetic customer request Inc4 showed that it was ambiguously formulated.

Overall, the qualitative and quantitative evaluations allow the conclusion that answering questions on the basis of historical data sets is also a promising use case for the automation of TCS with LLMs.

## 5. CONCLUSION

In summary, the results of the three prototypes show that there is great potential for the use of LLMs to support TCS in text correction, summarization and question answering. At the same time, it becomes apparent that in the case of more complex cognitive tasks, such as answering questions, the requirements for quality assurance of the results increase. Here, ambiguous question formulations quickly lead to unsatisfactory results. Full automation should therefore be avoided, especially in the case of business-critical customer requirements, because deviations in content and, in extreme cases, hallucinations in the responses of LLMs cannot be ruled out one hundred percent. However, our results show that the use of LLMs in so-called hybrid intelligence systems (Dellermann et al., 2019) can significantly increase productivity in TCS. SMEs should also actively deal with this topic, because our results show that in many application scenarios no in-house developments are necessary, but standard solutions can be used.

With our results, we contribute to the emerging theory about the potential and technical feasibility of LLM in service management. In addition, we offer concrete insights for operators of TCS units. We demonstrate the automation of cognitive tasks using real-world examples.

However, it is important to acknowledge the limitations of this work in order to classify this work and point out future research directions. The first limitation of this study is that it is based on technological prototypes developed with limited data. While these prototypes provide a proof of concept, they may not fully represent the complexity and variations of real-world scenarios. Further research is therefore needed, including large-scale validations of technological feasibility. The prototypes need to be tested on a larger scale to ensure they can handle the volume and variety of data in a real-world environment. This would provide a more robust validation of LLM capabilities and their operational readiness. Another limitation is the need to empirically investigate the usefulness of the technology in a technical customer service context. While the technology may show promise in a controlled environment, its effectiveness needs to be evaluated in a hands-on setting, such as a TCS department. This includes examining factors such as user acceptance, ease of integration into existing systems, and impact on service quality and efficiency. The study is domain-specific and focuses on the telecommunications sector. Therefore, the results may not be generalized to other customer service domains without further research. Each domain has its unique characteristics and challenges, and a solution that works in one domain does not necessarily lead to success in another. Cross-industry studies are needed to explore the applicability and adaptability of LLMs for TCS in different sectors. Finally, this work has contributed to the advancement of knowledge in the demonstration of the LLM potential for TCS, but the existing limitations underline the need for further research.